\documentclass{elsart}
\usepackage{graphicx}
\usepackage{amssymb}
\usepackage{amstext}
\usepackage{bm}

\def\be{\begin{equation}}
\def\ba{\begin{eqnarray}}
\def\ee#1{\label{#1}\end{equation}}
\def\ea#1{\label{#1}\end{eqnarray}}
\def\la{\langle}
\def\ra{\rangle}
\def\bs{\begin{center}}
\def\es{\end{center}}
\def\mv{\la v \ra}
\def\mvv{\la v^2 \ra}

\graphicspath{{img/}}

\begin{document}
\begin{frontmatter}

\title{
    Forcing inertial Brownian motors: efficiency and negative differential mobility
     }
\author[a]{Marcin Kostur}
  \author[a,b]{Lukasz Machura}
  \ead{Marcin.Kostur@physik.uni-augsburg.de} \author[a]{Peter
    H\"anggi} \author[b]{Jurek Luczka} \author[a]{Peter Talkner}
  \address[a]{Institute of Physics, University of Augsburg,
    Universit\"atsstrasse 1, D-86135 Augsburg, Germany}
  \address[b]{Institute of Physics, University of Silesia, P-40-007
    Katowice, Poland}

\begin{abstract}
  The noise-assisted, directed transport in a one-dimensional dissipative, inertial
  Brownian motor of the rocking type that is exposed to an external
  bias is investigated. We demonstrate that the velocity-load
  characteristics is distinctly non-monotonic, possessing regimes with
  a {\em negative differential mobility}.  In addition, we evaluate
  several possible efficiency quantifiers which are compared among
  each other. These quantifiers characterize the mutual interplay
  between the viscous drag and the external load differently, weighing
  the inherent rectification features from different physical
  perspectives.
\end{abstract}


\end{frontmatter}

\section{Introduction}
\label{intro} Brownian motors are small physical micro- or even
nano-machines that operate far from thermal equilibrium by
extracting the energy from both, thermal and nonequilibrium
fluctuations in order to generate work against external loads
\cite{HanBar1996,Astumian:Today,Reimann:Ratchets,ReiHan2002,Linke2002,HanAnn05}.
They present the physical analogue of bio-molecular motors that also
work out of equilibrium to direct intracellular transport and to
control motion and sensation in cells \cite{Julicher1997}.
The most popular models assume an overdamped Brownian dynamics
\cite{bartussek94,luczka1995,Hanepl1996,KosLuc2001,noriepl04,noripre04}.
In many situations, such as in biological applications, such a
simplification can be well justified from physical grounds.  There
exist several situations, however, where the inertial effects are
prominent \cite{pollak93,borromeo2000}; being intrinsically the case
for quantum Brownian motors \cite{reimann97,goychuk98}. In this
paper we will deal with inertial Brownian motors
\cite{jung96,mateos1,lindner:inertia,Sintes2002,Son2003,sengupta2004,machura1,machura2}.
The underlying deterministic dynamics can be chaotic
\cite{jung96,mateos1,sengupta2004,family1,barbi2000} and thus it is
distinctly more complex than its overdamped counterpart
\cite{bartussek94,Bartussek1996}. Despite an abundance of research
works dealing with numerous variants of Brownian motors and Brownian
ratchets
\cite{HanBar1996,Astumian:Today,Reimann:Ratchets,ReiHan2002,Linke2002,HanAnn05},
there remain still intriguing features awaiting to be discovered.
This present study is to the best of our knowledge the first work
that considers the behavior of the noise-activated, {\it directed
current} of an inertial Brownian motor versus an {\it external
bias}; thus yielding the {\it velocity-load behavior} when inertial
effects dominate. Here, we will demonstrate that a rocked, inertial
Brownian motor degree of freedom, if put to work against a load, can
exhibit {\it negative
  differential mobility}. This striking phenomenon has been observed
within a quantum mechanical setting for electron transfer phenomena
\cite{NavCan1976} or for ac-dc-driven tunnelling transport
\cite{Hartmann1997}, in the dynamics of cooperative Brownian motors
\cite{BroBen2000,BroCle2002,EicRei2004b}, Brownian transport with
complex topology (entropic ratchets)
\cite{White1984,Balakrishnan1995,CecMag1996,Slater1997,EicRei2002a,EicRei2002b}
and in some stylized, multistate models with state-dependent noise
\cite{CleBro2002,HalMan2004}, to name but a few.\\
Furthermore, we also investigate the efficiency for this forced
Brownian inertial transport; in this case, it is possible to devise
and to compare several qualifiers characterizing the efficiency of
energy conversion and of rectification.

\section{Biased, rocked inertial Brownian motor}
\label{model} Upon introducing an appropriate scaling of time and
length (the details are elaborated in Ref. \cite{machura1}) the
dynamics of a massive Brownian particle can be written in
dimensionless form; i.e.,
\begin{equation}
\ddot{{x}} + {\gamma} \dot{ x} =- {V}'({x}) + {F}
+ a \cos(\omega {t}) + \sqrt{2{\gamma}D} \; {\xi}
({t}), \label{NLbw}
\end{equation}
where $\gamma$ denotes the friction coefficient, $V(x)=V(x+1)$ is a
spatially periodic and asymmetric ratchet potential (i.e. no
reflection symmetry holds) with both, the period and the barrier
height set equal to one.  The quantity $F$ denotes the external,
constant load force. Additionally, the particle is driven by an
unbiased, time-periodic force of amplitude $a$ and angular frequency
$\omega$.  The interaction with the thermal bath is modeled by white
Gaussian noise $\xi(t)$ with auto-correlation function $\langle
\xi(t)\xi(s)\rangle = \delta(t-s)$, satisfying Einstein's
fluctuation-dissipation relation. $D$ stands for the re-scaled noise
intensity and as such it is proportional to the physical
temperature.

For the ratchet potential $V(x)$ we choose a linear superposition of
three spatial harmonics \cite{machura1}; i.e.,
\begin{eqnarray}\label{pot}
V(x) = V_0 [\sin(2 \pi x) + c_1 \sin (4 \pi x) + c_2 \sin (6 \pi
x)],
\end{eqnarray}
where $V_0$ normalizes the barrier height to unity and the parameters
$c_1$ and $c_2$ determine the specific ratchet profile.  Below, we
analyze in detail the case when $c_1=0.245$ and $c_2=0.04$, yielding
$V_0\simeq{0.461...}$.

\section{Rectification efficiency in presence of friction and load}
\label{sec:efficiency} The efficiency of a machine is defined as the
ratio of the power $P=F\mv$ done against en external force $F$ and
the input power $P_{in}$, i.e. $\eta = P/P_{in}$.  The same
definition of {\it efficiency of energy conversion} was used for
Brownian motors
\cite{Reimann:Ratchets,ReiHan2002,Linke2002,Zhou96,sekimoto97}:
\be \eta_{E} = \frac{F\mv}{P_{in}}.  \ee{EtaEnergy}
A grave disadvantage of such a characterization is that it yields a
vanishing measure (i.e. $\eta_{E} = 0$) in the absence of a load
force $F$. In many cases, however, like e.g. for protein transport
within a cell, the Brownian motor operates at a zero bias regime
($F=0$) and its objective is to carry a cargo across a viscous
environment. Clearly, the minimal energy input required to move a
particle in presence of friction $\gamma$ over a given distance
depends on the velocity, tending to zero when we move it very
slowly. If one is interested in delivering the cargo in a finite
time one should require that the transport is accomplished at an
average motor velocity $\mv$. In this case, the necessary energy
input is finite. Thus, we replace the load force in the expression
(\ref{EtaEnergy}) by the viscous force $\gamma\mv$ to obtain the
called {\it Stokes efficiency} \cite{wang}; i.e.,
\be \eta_{S} = \frac{\gamma \mv^2}{P_{in}}.  \ee{EtaStokes}
Upon combining the two above given notions we recover the {\it
  rectification efficiency} originally proposed by Suzuki and Munakata
\cite{munakata,munakata2005} or its equivalent version presented by
Derenyi {\it et al.} \cite{Derenyi1999}
\be \eta_{R} = \frac{F \mv + \gamma
  \mv^2}{P_{in}}.  \ee{EtaRec}
It is made up of the sum of the efficiencies $\eta_{S}$ and
$\eta_{E}$. Therefore, it accounts for both, the work that the
Brownian motor performs against the external bias $F$ as well as the
work that is necessary to move the object a given distance in a
viscous environment at the average velocity $\mv$.

The average input power for a tilted rocking ratchet is given by
\cite{machura1,machura2,Linke2005}: \be P_{in} = F \mv + \gamma [
\mvv - D_0 ].  \ee{Pin} This expression  follows from an energy
balance of the underlying equation of motion (\ref{NLbw})
\cite{machura1}.

\section{Numerical analysis }
\label{results}

Focussing on the directed current, we investigate the asymptotic,
time-periodic regime after effects of the initial conditions and
transient processes have died out. Then, the statistical quantifiers
of interest can be determined in terms of the statistical average over
the different realizations of the process (\ref{NLbw}) and over the
driving period $T$.

Clearly, there exist no analytical methods of analyzing eq.
(\ref{NLbw}) in presence of inertia. Therefore, we performed
extensive, precise numerical studies by employing the Stochastic
Runge-Kutta (SRK) algorithm of order 2 with a time step $h=10^{-3}$.
For the initial conditions we used a uniform distribution of the
initial position $x(t=t_0)$ at time $t_0$ on an interval lying between
two neighboring maxima of the ratchet potential given in (\ref{pot}).
The initial starting velocities $v(t=t_0)$ were randomly chosen from
an uniform distribution over the interval $[-0.2,0.2]$. All quantities
were averaged over 100 different trajectories, each of which evolved
over $10^5$ driving-periods $T$.  For the investigation of the
efficiency quantifiers defined in section \ref{sec:efficiency} above,
we restrict the discussion here to a set of optimal driving
parameters, reading, $a=3.7$, $D_0=0.001$, $\omega=4.9$ and
$\gamma=0.9$ (see for the details in Ref.  \cite{machura1,machura2}).

\subsection{Current-load behavior }
\label{num:DNM}
\begin{figure}[htbp]
\centerline{\includegraphics[angle=0,scale=1]{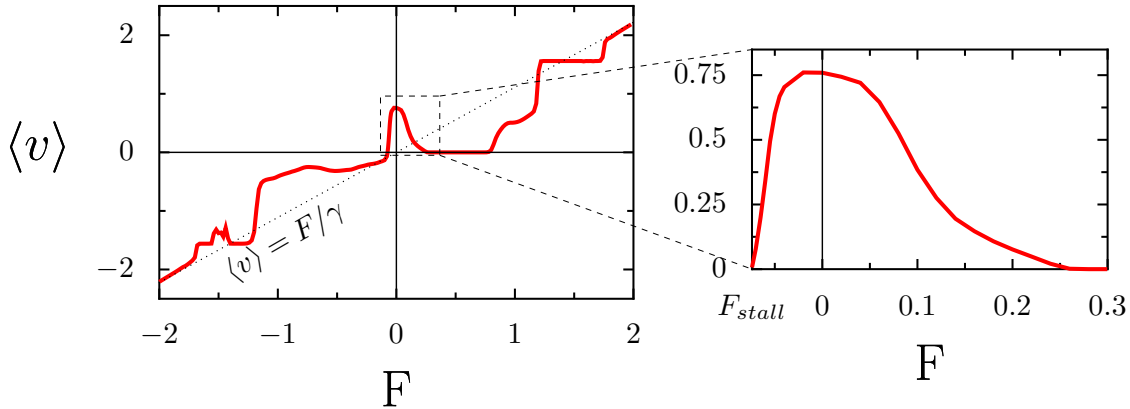}}
\caption{ %
  Average velocity of the inertial Brownian motor (\ref{NLbw}) as a
  function of an external, constant force $F$. The system parameters
  are: $a=3.7$, $\omega=4.9$, $\gamma=0.9$ and $D=0.001$.  The dotted
  line denotes the average velocity of a particle moving in the
  absence of a periodic potential, being the limiting case for the
  Brownian motor dynamics at $F\to\infty$. One can notice a few
  regimes where the differential mobility ($\partial \mv /\partial F$)
  assumes a negative value.  The most pronounced such behavior occurs
  for small positive values of the bias $F$ (depicted in the inset).
  For bias forces $F \in (F_{stall},0)$, $F_{stall} \simeq -0.074$,
  the Brownian motor performs against the external load.  }
\label{fig1}
\end{figure}

In Fig. \ref{fig1} we depict the load-velocity characteristics of the
non-equilibrium Brownian motor dynamics (\ref{NLbw}). Contrary to the
familiar, usually monotonic dependence found for overdamped ratchet
dynamics \cite{HanBar1996,Bartussek1996,Zapata96}, the
velocity-load-behavior becomes now considerably more complex,
exhibiting distinct non-monotonic characteristics. Around the forces
$F\simeq-1.4$ and $F\simeq{0}$ an increase of the bias $F$ results in
a corresponding decrease of the average velocity. This behavior is
termed {\it negative differential mobility}. The effect is extremely
pronounced in the neighborhood of $F=0$.
\label{mobility}
\\
Let us elucidate the underlying working mechanism in greater detail:
At a zero load the corresponding deterministic dynamics possesses one
stable attractor of period one (in velocity space, see in
\cite{machura2}) which translocates the particle from one to the next
potential well during one period $T$ of driving.  The particle moves
with a high Stokes efficiency as a consequence of small fluctuations
of the velocity from its average value. A residence within this
regime, however, requires that all system parameters are precisely
tuned. A consecutive increase of the external load $F$, regardless of
its sign, drives the system away from this most efficient regime and
the average velocity starts dropping to small values.  This is a
result of the complex inertial dynamics where a forcing of the
particle into the direction of its motion diminishes, rather than
increases the average velocity. In contrast, at very large magnitudes
of the load force $F$, the velocity assumes its its asymptotic value,
reading $\mv=F/\gamma$.

\subsection{Efficiency for forced, rocking Brownian motors}
\label{num:efficiency} 
\begin{figure}[htbp]
  \centerline{\includegraphics[angle=0,scale=1]{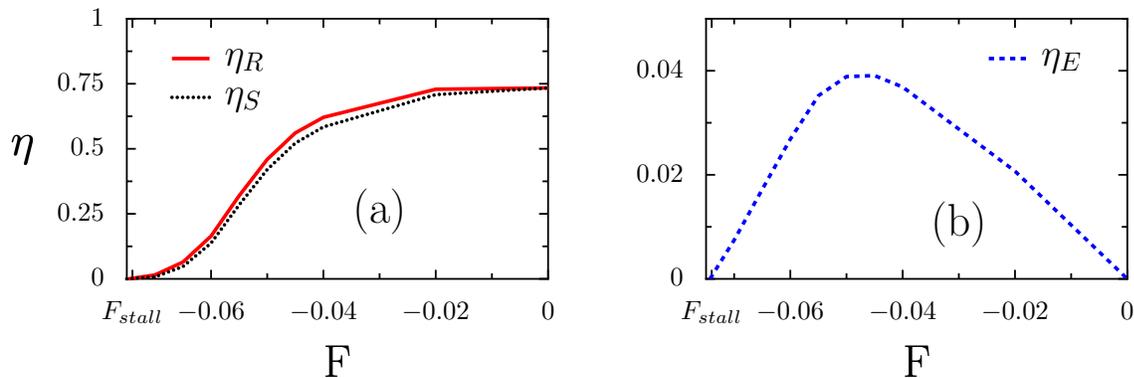}}
  \caption{Behavior of different efficiency measures within the regime
    of "uphill motion". Depicted are the efficiency of rectification
    $\eta_R$, the closely related Stokes efficiency $\eta_S$, in panel
    (a), and efficiency of energy conversion $\eta_E$, panel (b),
    versus the external load $F$, varying between the stall force
    $F_{stall}$ and the vanishing bias $F=0$.  The Stokes efficiency
    assumes much larger values than the corresponding energetic one;
    it is therefore dominating the viscous, noise-assisted transport.
  }
\label{fig2}
\end{figure}

As we remarked already above, near the bias
$F\simeq 0$, the Brownian motor operates optimally.  With Fig.
\ref{fig2} (a), we depict the behavior of the Stokes efficiency
within an interval of bias forces $F \in (F_{stall},0)$ where the
motor does work against the external force. The Stokes efficiency
assumes a value around $0.75$ at $F=0$, and monotonically decreases,
reaching zero at the stall force $F_{stall}$, where the average
velocity vanishes. In between, $F \in (F_{stall},0)$, the ratchet
device pumps particles uphill, cf. Fig. \ref{fig1}. The behavior of
the rectification efficiency in this regime closely matches the
behavior of the Stokes efficiency. Indeed, within this forcing
regime the efficiency of energy transduction $\eta_{E}$ assumes much
smaller values, see Fig. \ref{fig2} (b). Within this forcing regime
the bell-shaped character of $\eta_E$ is an immediate consequence of
its definition in eq. (\ref{EtaEnergy}): It acquires vanishing
values of both, at the stall force, where the velocity becomes zero
and at $F=0$, where the output power vanishes. In this regime
$P_{in}$ varies only slightly.

\section{Summary}
Biased, inertial rocking Brownian motors can exhibit an intriguing
velocity-load characteristics. We discovered that the average
velocity assumes a non-monotonic behavior as a function of the
external load; i.e., in certain regimes of external forcing the
differential mobility is {\it negative-valued}.  Near small negative
load forces the rocked, inertial Brownian motor is able to perform
``uphill''-motion against the external force. Within this regime the
bell-shaped energetic efficiency $\eta_{E}$ is distinctly smaller
than the corresponding efficiency of rectification and also smaller
than the related ``Stokes'' efficiency. These latter two
efficiencies clearly dominate over conversion of energy within this
very regime, where particles move against an externally applied
load.

\ack The authors gratefully acknowledge financial support by the
Deutsche Forschungsgemeinschaft via grant HA 1517/13-4, the
Graduiertenkolleg 283 (LM, PH, PT), the collaborative research grant
SFB 486 (PH), the DAAD-KBN (German-Polish project {\it Stochastic
  Complexity}) (JL,PH), and the ESF, Program {\it Stochastic Dynamics:
  fundamentals and applications; STOCHDYN} (JL,PH), and the Bavarian
collaboration program: Bavaria-Rio de Janeiro (PH).

\end{document}